\documentclass[conference, 9pt]{IEEEtran}
\IEEEoverridecommandlockouts
\usepackage{cite}
\usepackage{amsmath,amssymb,amsfonts}
\usepackage[noend]{algorithmic}
\usepackage{algorithm}
\usepackage{graphicx}
\usepackage{textcomp}
\usepackage{xcolor}
\def\BibTeX{{\rm B\kern-.05em{\sc i\kern-.025em b}\kern-.08em
		T\kern-.1667em\lower.7ex\hbox{E}\kern-.125emX}}
\usepackage{amsmath}
\usepackage{booktabs}
\usepackage{multirow}
\usepackage{xcolor}
\usepackage[super]{nth}
\usepackage{tikz}
\newcommand*\circled[1]{\tikz[baseline=(char.base)]{
		\node[shape=circle,draw,inner sep=0.2pt] (char) {#1};}}
\newcommand*\circledB[1]{\tikz[baseline=(char.base)]{
            \node[shape=circle,fill,inner sep=0.2pt] (char) {\textcolor{white}{#1}};}}
\usepackage{soul}
\usepackage{enumitem}

\usepackage{fancyhdr}
\fancypagestyle{firstpage}{
  \fancyhf{}
  \chead{To appear at the 58th IEEE/ACM Design Automation Conference (DAC), December 2021, San Francisco, CA, USA.}
  \cfoot{\thepage}
}

\usepackage[none]{hyphenat}

\begin{document}

\title{SparkXD: A Framework for Resilient and Energy-Efficient Spiking Neural Network Inference using Approximate DRAM
\vspace{-0.3cm}
}
	
\author{\IEEEauthorblockN{\textsuperscript{1}Rachmad Vidya Wicaksana Putra, \textsuperscript{2}Muhammad Abdullah Hanif, \textsuperscript{3}Muhammad Shafique}
\IEEEauthorblockA{\textsuperscript{1,2}\textit{Technische Universit\"at Wien (TU Wien)}, Vienna, Austria \\
\textit{\textsuperscript{3}New York University Abu Dhabi (NYUAD)}, Abu Dhabi, United Arab Emirates \\
Email: \{rachmad.putra, muhammad.hanif\}@tuwien.ac.at,
muhammad.shafique@nyu.edu}
\vspace{-0.9cm}
}

\maketitle
\thispagestyle{firstpage}

\begin{abstract}
Spiking Neural Networks (SNNs) have the potential for achieving low energy consumption due to their biologically sparse computation. 
Several studies have shown that the off-chip memory (DRAM) accesses are the most energy-consuming operations in SNN processing. 
However, state-of-the-art in SNN systems do not optimize the DRAM energy-per-access, thereby hindering achieving high energy-efficiency. 
To substantially minimize the DRAM energy-per-access, a key knob is to reduce the DRAM supply voltage but this may lead to DRAM errors (i.e., the so-called approximate DRAM). 
Towards this, we propose SparkXD, a novel framework that provides a comprehensive conjoint solution for resilient and energy-efficient SNN inference  using low-power DRAMs subjected to voltage-induced errors. 
The key mechanisms of SparkXD are: (1) improving the SNN error tolerance through fault-aware training that considers bit errors from approximate DRAM, (2) analyzing the error tolerance of the improved SNN model to find the maximum tolerable bit error rate (BER) that meets the targeted accuracy constraint, and (3) energy-efficient DRAM data mapping for the resilient SNN model that maps the weights in the appropriate DRAM location to minimize the DRAM access energy. 
Through these mechanisms, SparkXD mitigates the negative impact of DRAM (approximation) errors, and provides the required accuracy. 
The experimental results show that, for a target accuracy within 1\% of the baseline design (i.e., SNN without DRAM errors), SparkXD reduces the DRAM energy by ca. 40\% on average across different network sizes. 
\end{abstract}

\begin{IEEEkeywords}
Spiking neural networks, SNNs, inference, resilience, energy, efficiency, approximate computing, DRAM, DRAM errors, error-tolerance. 
\vspace{-0.3cm}
\end{IEEEkeywords}

\section{Introduction}

\begin{figure}[t]
\centering
\includegraphics[width=\linewidth]{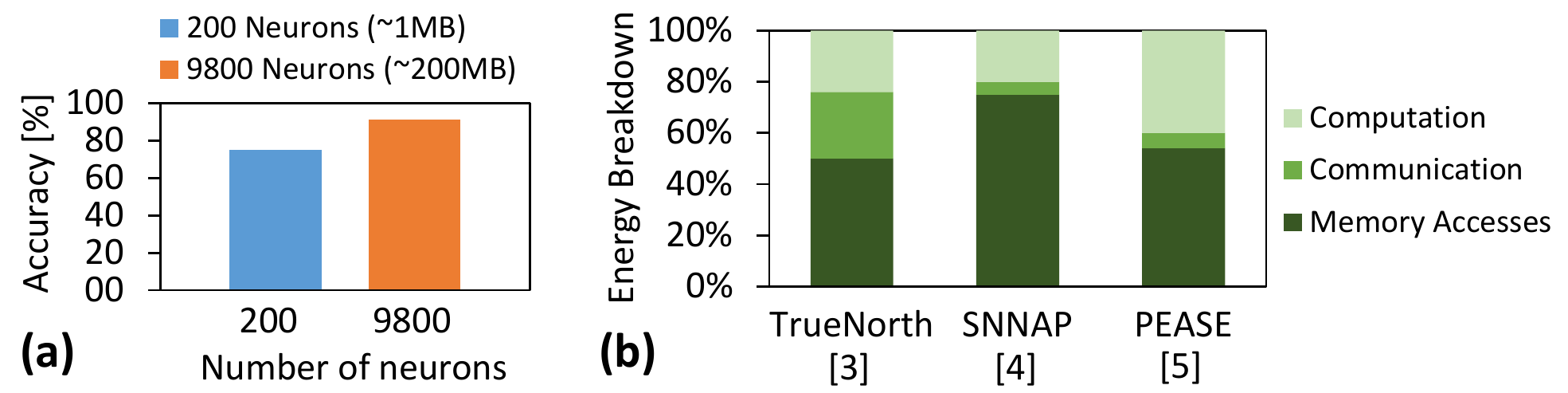}
\vspace{-0.7cm}
\caption{(a) Accuracy achieved by small- and large-sized SNN models on MNIST dataset. (b) Energy breakdown of SNN processing on different hardware platforms, i.e., TrueNorth \cite{Ref_Akopyan_TrueNorth_TCAD15}, PEASE \cite{Ref_Roy_PEASE_ISLPED17}, and SNNAP \cite{Ref_Sen_ApproxSNN_DATE17} (adapted from the studies presented in \cite{Ref_Krithivasan_SpikeBundle_ISLPED19}).}
\label{Fig_ObserveEnergyBreak}
\vspace{-0.6cm}
\end{figure}

Spiking neural networks (SNNs) bear the potential of achieving low energy processing and high algorithmic performance because of their biological plausibility \cite{Ref_Pfeiffer_DLSNN_FNINS18}.
A large-sized SNN model is more desirable as it can achieve higher accuracy than the smaller ones. 
For instance, Fig.~\ref{Fig_ObserveEnergyBreak}(a) shows that the 200MB-sized SNN model achieves 92\% accuracy for MNIST, while the 1MB-sized SNN achieves only 75\%. 
On the other hand, most of the SNN hardwares employ an on-chip memory of limited size, e.g., less than 100MB \cite{Ref_Akopyan_TrueNorth_TCAD15,Ref_Roy_PEASE_ISLPED17,Ref_Sen_ApproxSNN_DATE17}. 
Therefore, running an SNN model, whose size is larger than the on-chip memory, on such a hardware needs excessive DRAM accesses. 
This problem is even more critical for Edge-AI applications where SNNs are deployed on low-cost embedded devices with a small-sized on-chip memory, leading to high DRAM energy consumption. 
This hinders (embedded) SNN hardwares from achieving further energy-efficiency gains, as the DRAM access energy is relatively higher than other operations (e.g., neuron computations). 
Previous work \cite{Ref_Krithivasan_SpikeBundle_ISLPED19} observed that the memory access energy during inference is dominant, consuming ca. 50\%-75\% of the total energy across different hardware platforms, as also shown in Fig.~\ref{Fig_ObserveEnergyBreak}(b).
\textit{The challenging research question is:}
\textit{If and how can we reduce the DRAM access energy for the SNN inference, while maintaining the accuracy.} 
The solution to this issue will enable efficient SNN inference for energy-constrained devices and their applications for IoT-Edge and Smart CPS (cyber physical systems). 

\subsection{State-of-the-Art and Their Limitations}
\label{Sec_SOTA}

To reduce the energy of SNN inference, the state-of-the-art works have proposed different techniques, and can be categorized as follows:
\begin{itemize}[leftmargin=*]
    \item \textbf{Reduction of the SNN operations} through weight pruning \cite{Ref_Rathi_PruneQuantizeSNN_TCAD18}, stochastic neural operations \cite{Ref_Sen_ApproxSNN_DATE17}, and neuron elimination \cite{Ref_Putra_FSpiNN_TCAD20}. 
    These approaches can reduce the number of DRAM accesses required for SNN parameters. 
    \item \textbf{Quantization} that reduces the possible representable values for a single weight \cite{Ref_Rathi_PruneQuantizeSNN_TCAD18}\cite{Ref_Putra_FSpiNN_TCAD20}. 
    This approach decreases the amount of data (i.e., weights) to be stored in and fetched from DRAM.
\end{itemize}

\textbf{Limitations:} 
These state-of-the-art works mainly target reducing the number of accesses, but do not optimize the DRAM energy-per- access and approximations in DRAM that provide an additional knob to achieve high energy efficiency. 
Therefore, these works can potentially hinder the SNN inference system exploiting the full potential of DRAM energy saving. The goal should be jointly minimizing the energy-per-access as well as the number of accesses, leveraging approximations in DRAM to expose the full energy saving potential, while overcoming the adverse effects of the approximation-induced errors

To address these limitations, \textit{we propose to employ approximate DRAM (i.e., DRAM with reduced supply voltage) with efficient DRAM mapping and error-tolerant SNN training to substantially reduce the DRAM energy in SNN inference systems while preserving their accuracy}. 
Note, our approach can also be combined with the above-discussed state-of-the-art existing techniques to further improve the energy-efficiency of SNN inference. 
For instance, Fig.~\ref{Fig_CaseStudy}(a) shows the estimated DRAM energy benefits achieved by our proposed technique combined with the weight pruning.

To highlight the potential of approximate DRAM, we present an experimental case study in the following section.

\subsection{Motivational Case Study and Key Challenges}
\label{Sec_MotivationChallenges}

We analyze (1) the DRAM access energy consumed by different DRAM access conditions (i.e., \textit{a row buffer hit}, \textit{a row buffer miss}, and \textit{a row buffer conflict}), and (2) the dynamics of DRAM array voltage ($V_{array}$), for both the original and the approximate DRAM scenarios. 
In a row buffer hit, the requested data is already in the DRAM row buffer, and hence the data can be accessed directly.
Meanwhile, a row buffer miss or conflict has to open the requested DRAM row before the data can be accessed.
The detailed information of the different DRAM access conditions will be discussed in Section~\ref{Sec_Prelim_ApproxDRAM_Basic}.

\begin{figure}[t]
\centering
\includegraphics[width=\linewidth]{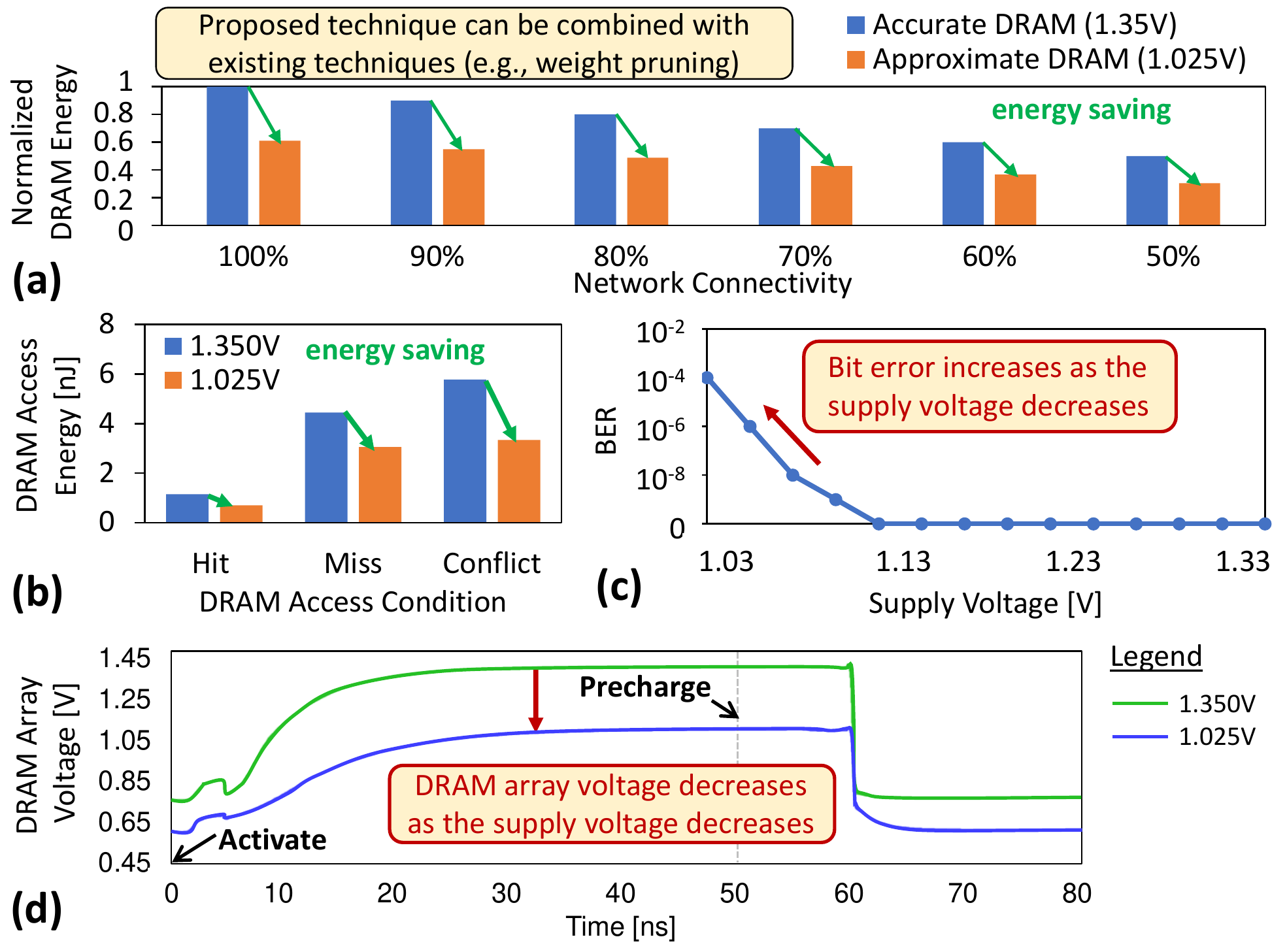}
\vspace{-0.7cm}
\caption{(a) The estimated DRAM energy benefits achieved by our proposed technique combined with the weight pruning, across different rates of network connectivity (synaptic weight connections) for a network with 4900 neurons. 
The results are obtained from experiments using DRAMPower \cite{Ref_DRAMpower} and a LPDDR3-1600 4Gb DRAM configuration.
(b) DRAM access energy for different DRAM access conditions (i.e., a row buffer hit, a row buffer miss, and a row buffer conflict). 
(c) Bit error rate (BER) and the corresponding DRAM supply voltage values (based on the study of \cite{Ref_Chang_Voltron_POMACS17}). (d) The dynamics of DRAM array voltage under different supply voltage values.}
\label{Fig_CaseStudy}
\vspace{-0.5cm}
\end{figure}

For the experimental setup, we use the LPDDR3-1600 4Gb DRAM configuration. 
We employ the \textit{DRAMPower} simulator \cite{Ref_DRAMpower} to obtain the DRAM access energy, as it is widely used in memory and architecture communities and has been validated against real measurements \cite{Ref_Chandrasekar_DRAMPower}.
We also employ the DRAM circuit model from \cite{Ref_Chang_Voltron_POMACS17} and the SPICE simulator to study the dynamics of DRAM array voltage.
The original DRAM is operated with 1.35V supply voltage, while the approximate one with 1.025V. 
More details on the experimental setup are presented in Section~\ref{Sec_EvalMethod}.
From the experimental results presented in Figs.~\ref{Fig_CaseStudy}(b) and \ref{Fig_CaseStudy}(d), we make the following key observations.
\begin{itemize}[leftmargin=*]
    \item The reduced DRAM voltage reduces the DRAM energy-per-access, across different DRAM access conditions, i.e., 31\%-42\% energy savings per access.
    \item The row buffer hit incurs the least energy consumption, compared to the row buffer miss and the row buffer conflict conditions. 
    Therefore, the row buffer hit should be maximized to reduce the DRAM access energy.
\end{itemize}

Although employing the approximate DRAM can significantly reduce the DRAM energy-per-access, it reduces the DRAM reliability as the bit errors increase along with the decreased supply voltage (see Fig.~\ref{Fig_CaseStudy}(c)). 
These errors will consequently reduce the accuracy of the SNN as they alter the weight values in DRAM. 
Therefore, \textit{the key challenge is how to ensure high energy savings considering approximate DRAMs, while minimizing their negative impact on the SNN inference accuracy under the targeted accuracy constraints.}

\vspace{-0.1cm}
\subsection{Our Novel Contributions}
\label{Sec_Novelty}
\vspace{-0.1cm}

\begin{figure}[hbtp]
\vspace{-0.3cm}
\centering
\includegraphics[width=0.95\linewidth]{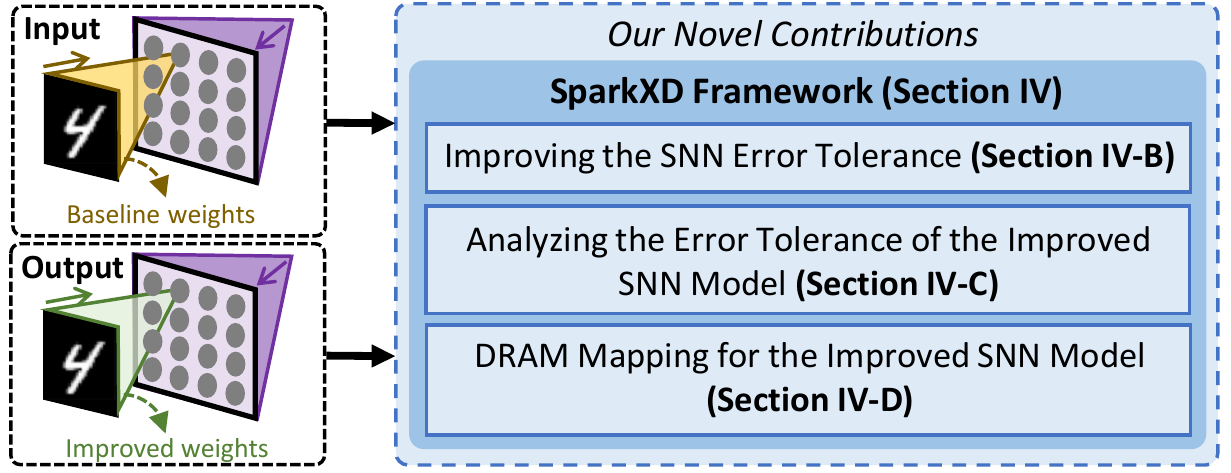}
\vspace{-0.3cm}
\caption{An overview of our novel contributions (highlighted in the blue boxes).}
\label{Fig_NovelContributions}
\vspace{-0.3cm}
\end{figure}

To overcome the above key challenges, we propose \textbf{SparkXD framework} that enables a resilient and energy-efficient \underline{Sp}iking neur\underline{a}l netwo\underline{rk} inference using appro\underline{X}imate \underline{D}RAM.
To the best of our knowledge, this work is the first effort that exploits approximate DRAM for improving the energy-efficiency of SNNs while enabling their error-tolerant training.
Following are the novel steps performed in the SparkXD framework (the overview is illustrated in Fig.~\ref{Fig_NovelContributions}): 
\begin{enumerate}[leftmargin=*] 
  \item \textbf{Improving the SNN Error Tolerance}, so that the SNN inference achieves high accuracy, even in the presence of bit errors in approximate DRAMs. It is done by incorporating the error profiles from the given approximate DRAM into the training.
  \item \textbf{Analyzing the Error Tolerance of the Improved SNN Model} to find the maximum tolerable BER that can be applied to the SNN model, while meeting the targeted accuracy. 
  It is done by adjusting the BER values, while checking whether the obtained accuracy meets the user-specified accuracy.
  \item \textbf{DRAM Mapping for the Improved SNN Model} to maximize DRAM row buffer hit and optimize the DRAM access energy, by placing the synaptic weights into the appropriate DRAM partition (e.g., subarray) whose error profile meets the BER requirement.    
\end{enumerate}

\textbf{Key Results:} 
We evaluated SparkXD framework for (1) accuracy, using Python-based simulations on GPGPU and Embedded GPU, with the MNIST and the Fashion MNIST datasets\footnote{The research for the unsupervised learning-based
SNNs is still in early stage and mostly use small datasets like MNIST and
Fashion MNIST. We adopt the same test conditions as used widely by the
SNN research community.}; and (2) DRAM access energy, using DRAMPower \cite{Ref_DRAMpower}.
The experimental results show that, for a target accuracy within 1\% of the baseline SNN system with accurate DRAM, SparkXD reduces the DRAM access energy by approximately 40\% on average, across different network sizes.

\renewcommand{\headrulewidth}{0pt}
\vspace{-0.1cm}
\section{Background and Preliminaries}
\label{Sec_Prelim}

\subsection{Spiking Neural Networks (SNNs)}
\label{Sec_Prelim_SNNs}

\begin{figure}[hbtp]
\centering
\includegraphics[width=0.85\linewidth]{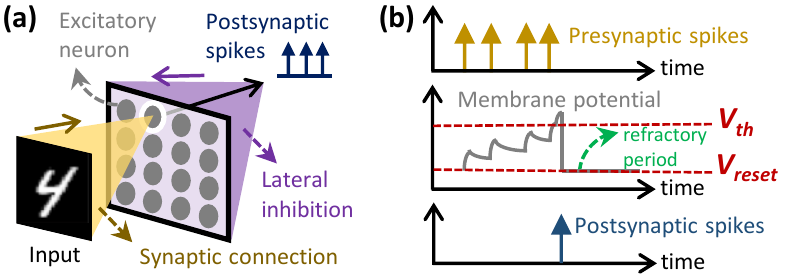}
\vspace{-0.3cm}
\caption{(a) The SNN architecture considered in this work. (b) The neuronal dynamics in LIF neuron model.}
\label{Fig_SNN_n_LIF}
\vspace{-0.5cm}
\end{figure}

SNNs are considered as the $3^{rd}$ generation of neural network models, as they are highly bio-plausible (using spikes for conveying information). 
The components of an SNN model include \textit{network architecture}, \textit{neuron and synapse models}, \textit{spike coding}, and the \textit{learning rule}.
In this work, we consider the SNN architecture of Fig.~\ref{Fig_SNN_n_LIF}(a), as it is the state-of-the-art for the unsupervised settings \cite{Ref_Putra_FSpiNN_TCAD20}. 
Each input pixel is connected to all excitatory neurons. 
Furthermore, each spike from each neuron is passed to other neurons to provide inhibition which promotes competition among neurons. 
For the neuron model, the Leaky Integrate-and-Fire (LIF) is used, due to its low complexity.  
Its membrane potential increases when a presynaptic spike comes, and otherwise, it decreases exponentially. 
If the potential reaches the threshold ($V_{th}$), a postsynaptic spike is fired, and then it goes back to the reset potential ($V_{reset}$), as shown in Fig.~\ref{Fig_SNN_n_LIF}(b). 
Meanwhile, the synapse is modeled by the synaptic conductance, which increases by weight ($w$) when a presynaptic spike arrives at a synapse, and otherwise, decreases exponentially.
To convert the input into a sequence of spikes (spike train), a spike coding is employed. There are different techniques of spike coding, e.g., rate, rank-order, phase, and burst \cite{Ref_Gautrais_SpikeCoding_Bio98, Ref_Thorpe_RankOrder_Springer98, Ref_Park_BurstSNN_DAC19}. 
For the learning rule, we consider the spike-timing-dependent plasticity (STDP), since it has been widely used by previous works. 

\subsection{Approximate DRAM}
\label{Sec_Prelim_ApproxDRAM}

\subsubsection{DRAM Fundamentals}
\label{Sec_Prelim_ApproxDRAM_Basic}

\begin{figure}[hbtp]
\vspace{-0.3cm}
\centering
\includegraphics[width=0.95\linewidth]{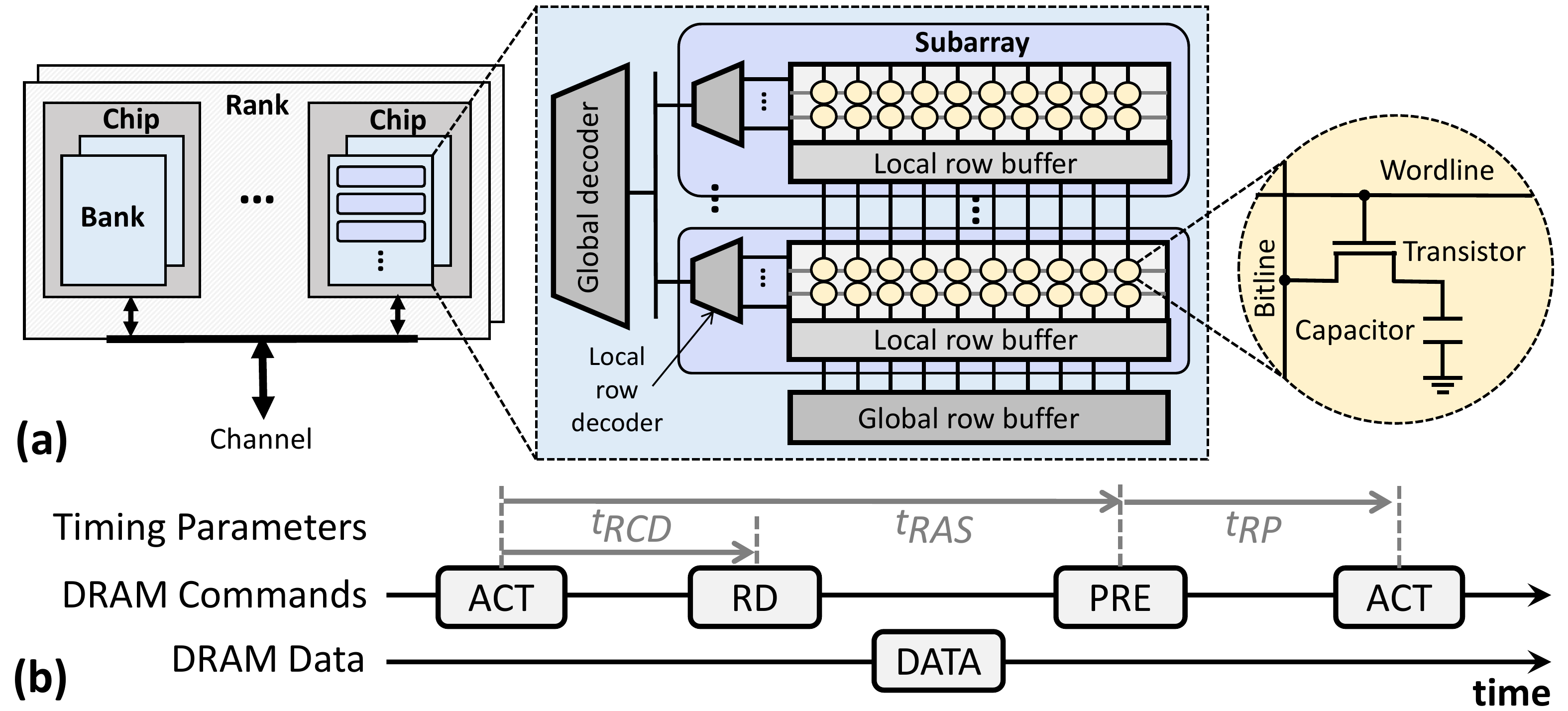}
\vspace{-0.3cm}
\caption{(a) Organization of a commodity DRAM. (b) Diagram of the DRAM commands and the DRAM timing parameters.}
\label{Fig_DRAM_OrgOps}
\vspace{-0.2cm}
\end{figure}

The organization of a DRAM consists of \textit{channel}, \textit{rank}, \textit{chip}, \textit{bank}, \textit{subarray}, \textit{row}, and \textit{column} \cite{Ref_Putra_DRMap_DAC20}, as shown in Fig. \ref{Fig_DRAM_OrgOps}(a). 
To operate the DRAM, several operations are involved. 
A single DRAM request can access data in parallel from multiple DRAM chips in the same rank. 
In each chip, the request is passed to a specific bank and decoded into the row and column addresses. 
Data from the requested row are copied to the row buffer when the \textit{activation} (ACT) command is issued. 
Then, data can be read from or written to a specific column in the activated row buffer when the \textit{read} (RD) or \textit{write} (WR) command  is issued. 
There are different possible DRAM access conditions: \textit{a row buffer hit}, \textit{a row buffer miss}, and \textit{a row buffer conflict} \cite{Ref_Putra_DRMap_DAC20}. 
A row buffer hit happens if the requested row is already activated, and the requested data is already in the row buffer. 
Hence, the data can be accessed directly. 
If the requested row is not yet activated, then the condition is either a row buffer miss or conflict. 
A row buffer miss happens if there is no activated row, hence it needs to activate the requested row before accessing the data. 
A row buffer conflict happens if the requested row is not yet activated, but the row buffer is still occupied by another activated row.
Hence, this condition needs to close the activated row using the \textit{precharging} (PRE) command, before activating the requested row using the \textit{activation} (ACT) command.
Fig.~\ref{Fig_DRAM_OrgOps}(b) shows the DRAM commands (i.e., ACT, RD or WR, and PRE) and the corresponding timing parameters (i.e., $t_{RCD}$: row address to column address delay, $t_{RAS}$: row active time, and $t_{RP}$: row precharge time). 

\subsubsection{Reduction of the DRAM Supply Voltage}
\label{Sec_Prelim_ApproxDRAM_ReducedV}

\begin{figure}[hbtp]
\centering
\includegraphics[width=0.9\linewidth]{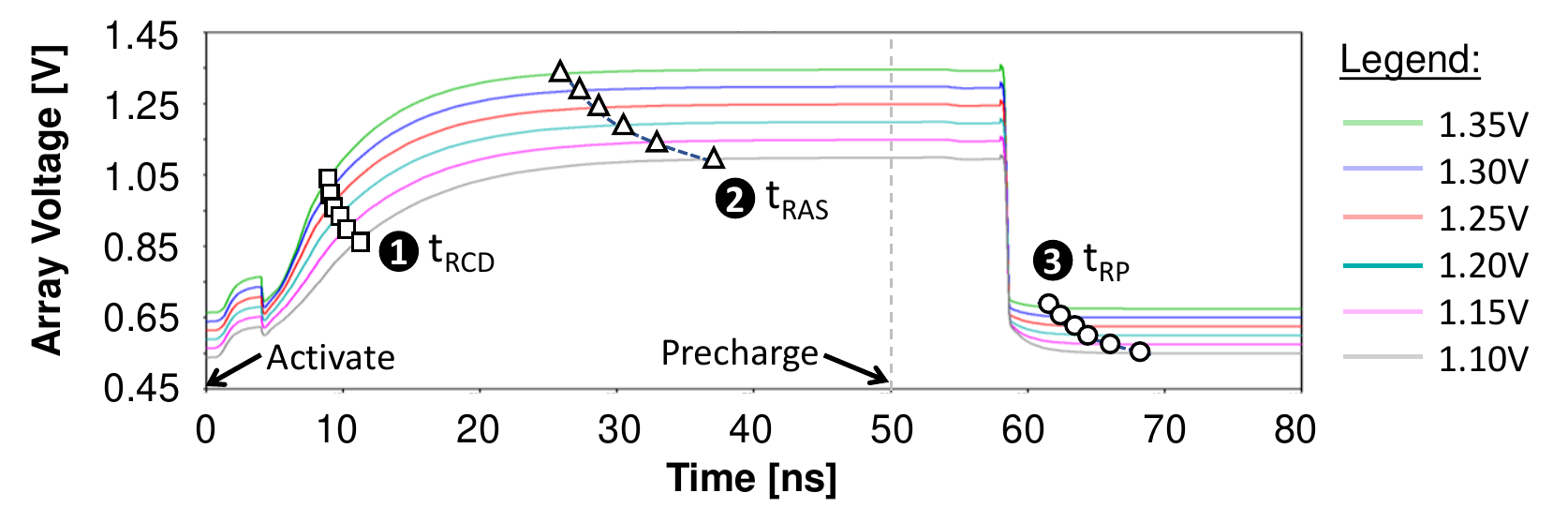}
\vspace{-0.3cm}
\caption{The dynamics of the DRAM $V_{array}$ and the timing parameters.} 
\label{Fig_SpiceDRAM}
\vspace{-0.5cm}
\end{figure}

We performed detailed experiments using SPICE and a DRAM circuit model from \cite{Ref_Chang_Voltron_POMACS17} to further characterize the DRAM parameters, e.g., the array voltage ($V_{array}$) and the timing parameters ($t_{RCD}$, $t_{RAS}$, and $t_{RP}$), under different supply voltage ($V_{supply}$) values. 
The results are shown in Fig.~\ref{Fig_SpiceDRAM}. These parameters are then used for DRAM energy estimation. 
The \textit{ready-to-access} voltage is obtained when the $V_{array}$ reaches 75\% of the $V_{supply}$.
It represents the minimum $t_{RCD}$ for reliable DRAM operations (see Label-\circledB{1}).
The \textit{ready-to-precharge} voltage is obtained when the $V_{array}$ reaches 98\% of $V_{supply}$.
It represents the minimum $t_{RAS}$ for reliable DRAM operations (see Label-\circledB{2}). 
The \textit{ready-to-activate} voltage is obtained when the $V_{array}$ is within 2\% of $V_{supply}/2$.
It represents the minimum $t_{RP}$ for reliable DRAM operations (see Label-\circledB{3}).

\section{Error Modeling for Approximate DRAM}
\label{Sec_ErrorModelDRAM}

Previous work in \cite{Ref_Koppula_EDEN_MICRO19} has proposed four probabilistic error models that closely fit the real approximate DRAM. 
\begin{itemize}[leftmargin=*]
    \item \textbf{Error Model-0:} The bit errors have a \textit{uniform random distribution} across a DRAM bank. 
    The errors are modeled by considering (1) the \textit{weak cells} (i.e., cells that fail when the DRAM parameters reduced), and (2) the probability of an error in any weak cell.
    \item \textbf{Error Model-1:} The bit errors have a \textit{vertical distribution} across the bitlines of a DRAM bank.
    The errors are modeled by considering (1) the weak cells in bitline $B$, and (2) the probability of an error in the weak cells of bitline $B$.
    \item \textbf{Error Model-2:} The bit errors have a \textit{horizontal distribution} across the wordlines of a DRAM bank.
    The errors are modeled by considering (1) the weak cells in wordline $W$, and (2) the probability of an error in the weak cells of bitline $W$.
    \item \textbf{Error Model-3:} It is a data-dependent error model, i.e., profile of the bit errors follow a \textit{uniform random distribution that depends on the content of the DRAM cells}.
    The errors are modeled by considering (1) the weak cells, (2) the probability of an error in the weak cells that contain a 1 value, and (3) the probability of an error in the weak cells that contain a 0 value.
\end{itemize}

In this work, we employ the DRAM Error Model-0, because: (1) it produces the errors with high similarity to the real approximate DRAM; (2) it provides a reasonable approximation of other error models, i.e., approximation of (a) error distribution across bitlines like Error Model-1, (b) error distribution across wordlines like Error Model-2, and (c) uniform random distribution like Error Model-3; and (3) it offers fast error injection by software. 
Previous work \cite{Ref_Koppula_EDEN_MICRO19} also used the Error Model-0 majorly due to the similar reasons.

\section{SparkXD Framework}
\label{Sec_SparkXD}

\vspace{-0.2cm}
\subsection{Overview}
\label{Sec_SparkXD_Overview}
\vspace{-0.1cm}

\begin{figure}[t]
\centering
\includegraphics[width=\linewidth]{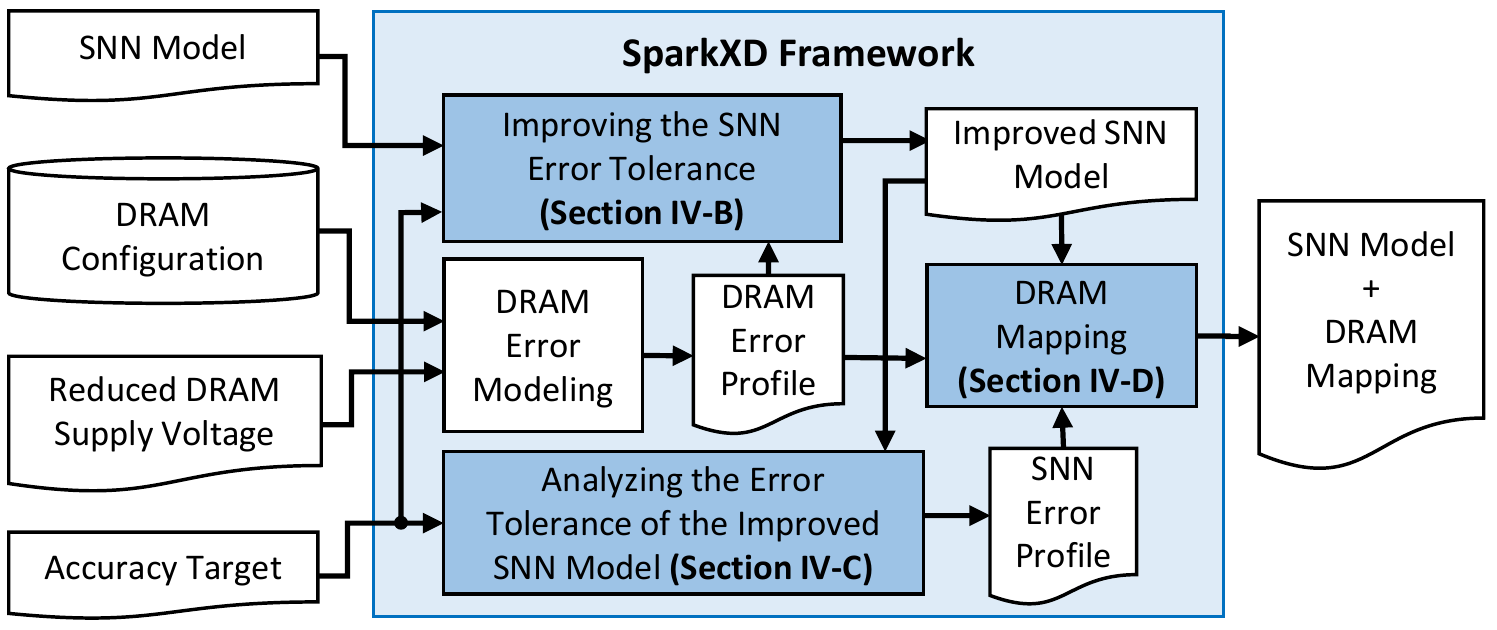}
\vspace{-0.7cm}
\caption{The SparkXD framework with novel steps highlighted in blue boxes.}
\label{Fig_SparkXD}
\vspace{-0.1cm}
\end{figure}

We propose the SparkXD framework to enable a resilient and energy-efficient SNN inference in the presence of voltage-induced DRAM errors. 
The detailed steps of SparkXD are shown in Fig.~\ref{Fig_SparkXD}, which are explained in the subsequent sections.
\begin{enumerate}[leftmargin=*]
  \item \textbf{Improving the Error Tolerance of SNN Model (Section~\ref{Sec_SparkXD_ImproveSNN})}. 
  It makes the SNN inference achieves high accuracy, even in the presence of bit errors in DRAM. It is performed by incrementally increasing the BER from 0 to a defined maximum BER in the training process, while considering the error profile from the DRAM error model.
  \item \textbf{Analyzing the Improved SNN Error Tolerance  (Section~\ref{Sec_SparkXD_AnalyzeSNN})}. 
  The idea is to find the highest BER for the improved SNN model that fulfills the user-specified inference accuracy. It employs a linear search on the given BER values, while checking if the corresponding accuracy meets the targeted accuracy. If so, the associated BER value is selected as the solution candidate.
  \item \textbf{DRAM Mapping for the Improved SNN (Section~\ref{Sec_SparkXD_DRAMmapping})}. It aims at optimizing the DRAM access energy for the improved SNN  model through the following ideas: 
  (1) data are mapped in the appropriate DRAM partition (e.g., subarray), whose error rate meets the BER requirement; and (2) data are mapped in a way that maximizes the row buffer hits, while exploiting DRAM multi-bank burst feature.
\end{enumerate} 

\subsection{Improving the SNN Error Tolerance}
\label{Sec_SparkXD_ImproveSNN}

The bit errors in the SNN weights can degrade the accuracy, since they alter the weight values and deviate the classification to a different class. 
Towards this, \textit{we improve the SNN error tolerance through a training process that incorporates the error profile of the approximate DRAM.} 
The proposed training has the following key steps. 

\begin{itemize}[leftmargin=*]
    \item \textbf{Step-1:} The bit errors are generated for different BER values, based on the different $V_{supply}$ values and the DRAM error model-0 that follows a uniform random distribution across a DRAM bank.
    \item \textbf{Step-2:} The generated bit errors are then injected into the locations in DRAM. 
    Therefore, the bits of data (weights) stored in these locations will be flipped.
    Here, as the baseline DRAM mapping, the weights are mapped in subsequent address space in a DRAM bank to exploit the DRAM burst feature and achieve high throughput. 
    If a DRAM bank is already filled, then the weights are mapped in the different banks of the same DRAM chip. 
    \item \textbf{Step-3:} Afterwards, we include the bit errors in the training process by incrementally increasing the BER value from a minimum error rate to a maximum one (the error rates are defined in the \textbf{Step-1}, while the locations of bit errors are defined in \textbf{Step-2}).
    Furthermore, we increase the BER after each epoch of training by a user-defined increment value (e.g., the next error rate is 10x of the previous one).
    In this manner, the SNN is gradually trained to tolerate the bit errors from the lowest rate to the given maximum rate, thereby carefully improving the SNN error tolerance.
\end{itemize}

\vspace{-0.3cm}
\subsection{Analyzing the Error Tolerance of the Improved SNN Model}
\label{Sec_SparkXD_AnalyzeSNN}
\vspace{-0.1cm}

\begin{figure}[hbtp]
\centering
\includegraphics[width=\linewidth]{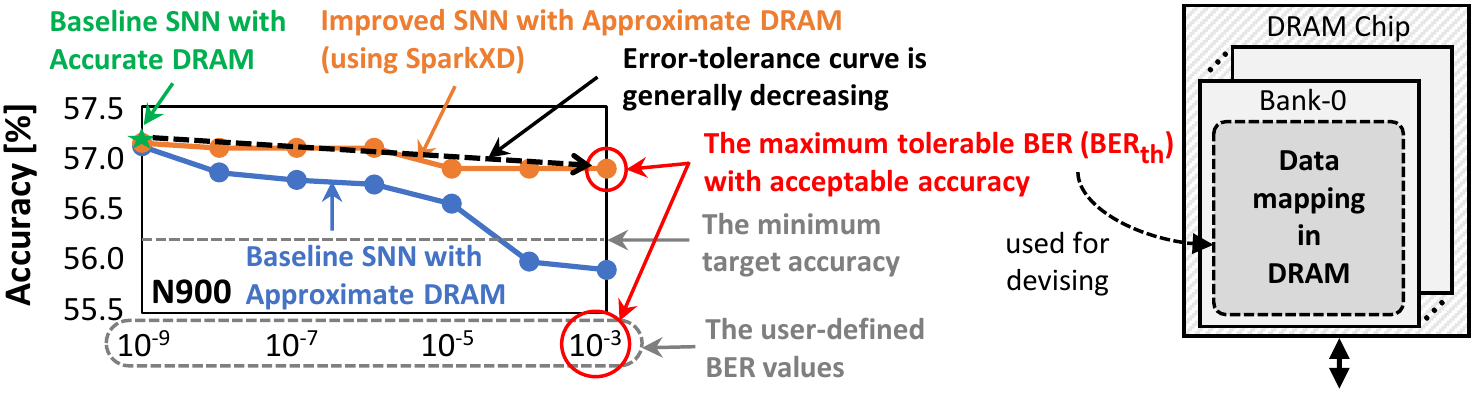}
\vspace{-0.7cm}
\caption{The error tolerance analysis finds the maximum tolerable BER of a given network (e.g., N900: a 900-neurons SNN) for devising DRAM mapping.}
\label{Fig_ErrorTolerance}
\end{figure}

The accuracy of SNN inference needs to be maintained within the user-specified accuracy, even in the presence of bit errors in the synaptic weights. 
Towards this, \textit{our SparkXD framework analyzes the error tolerance of the improved SNN model, to find the maximum tolerable BER that can be applied in the SNN systems.} 
It performs a linear search on the different BER values, which are obtained from Section~\ref{Sec_SparkXD_ImproveSNN}). 
It searches from a minimum error rate to a maximum one, while checking whether the corresponding accuracy meets the user-specified accuracy. 
The linear search can be employed, because we found that the SNN error-tolerance curve is generally decreasing as the BER increases (see Fig.~\ref{Fig_ErrorTolerance}). 
This technique ensures that bit errors whose rate is below or equal to the maximum tolerable BER, will not decrease the  accuracy under the user-defined target. 
This technique is also beneficial for devising a resilient and energy-efficient DRAM mapping, which is discussed in Section~\ref{Sec_SparkXD_DRAMmapping}.

To synergistically employ the techniques from Section~\ref{Sec_SparkXD_ImproveSNN} and Section~\ref{Sec_SparkXD_AnalyzeSNN}, we devise  Algorithm~\ref{Alg_TrainAnalyzeSNN} for improving and analyzing the SNN error tolerance. 

\begin{algorithm}[hbtp]
\scriptsize
\caption{\color{black} Improving and analyzing the SNN error tolerance}
\label{Alg_TrainAnalyzeSNN}
\begin{algorithmic}[1]
\renewcommand{\algorithmicrequire}{\textbf{INPUT:}}
\renewcommand{\algorithmicensure}{\textbf{OUTPUT:}}
\REQUIRE \textbf{(1)} Baseline SNN: model ($model_0$), accuracy ($model_0.acc$); 
\textbf{(2)} DRAM error model ($DRAMerr$); 
\textbf{(3)} BER: error rates ($rates$), number of error rates ($N_{rates}$); 
\textbf{(4)} Training dataset: samples ($S_{train}$), number of samples ($N_{train}$), number of epoch ($N_{epoch}$); 
\textbf{(5)} Test dataset: samples ($S_{test}$), number of samples ($N_{test}$); 
\textbf{(6)} Target accuracy: lower bound ($acc\_bound$); \\
\ENSURE \textbf{(1)} Improved SNN: model ($model_1$), accuracy ($model_1.acc$); \\ 
\textbf{(2)} Maximum tolerable BER ($BER_{th}$); \\
\vspace{0.1cm}
\renewcommand{\algorithmicrequire}{\textbf{BEGIN}}
\renewcommand{\algorithmicensure}{\textbf{END}}
\REQUIRE \hspace{0.1cm} \\   
    \textbf{Initialization}: \\
     \STATE $model_{temp} = model_0$; \\
    \textbf{Process}: \\
      \FOR{$i = 0$ to $(N_{rates}-1)$}
        \STATE $error = DRAMerr(rates[i])$; // error generation 
        \STATE inject $error$ into $model_{temp}$; // error injection 
        \FOR{$p = 0$ to $(N_{epoch}-1)$} 
          \FOR{$r = 0$ to $(N_{train}-1)$}
            \STATE train $model_{temp}$ with $S_{train}[r]$; // train
          \ENDFOR
        \ENDFOR
        \FOR{$s = 0$ to $(N_{test}-1)$}
          \STATE test $model_{temp}$ with $S_{test}[s]$; // test
        \ENDFOR
        \IF{$model_t.acc \geq (model_0.acc-acc\_bound)$}
          \STATE $model_1 = model_{temp}$;
          \STATE $model_1.acc = model_{temp}.acc$;
          \STATE $BER_{th} = rates[i]$;
        \ENDIF
      \ENDFOR
    \RETURN $model_1$;
\ENSURE 
\end{algorithmic}
\end{algorithm}
\setlength{\textfloatsep}{4pt}

\vspace{-0.2cm}
\subsection{DRAM Mapping for the Improved SNN Model}
\label{Sec_SparkXD_DRAMmapping}
\vspace{-0.1cm}

\begin{figure}[hbtp]
\vspace{-0.3cm}
\centering
\includegraphics[width=0.9\linewidth]{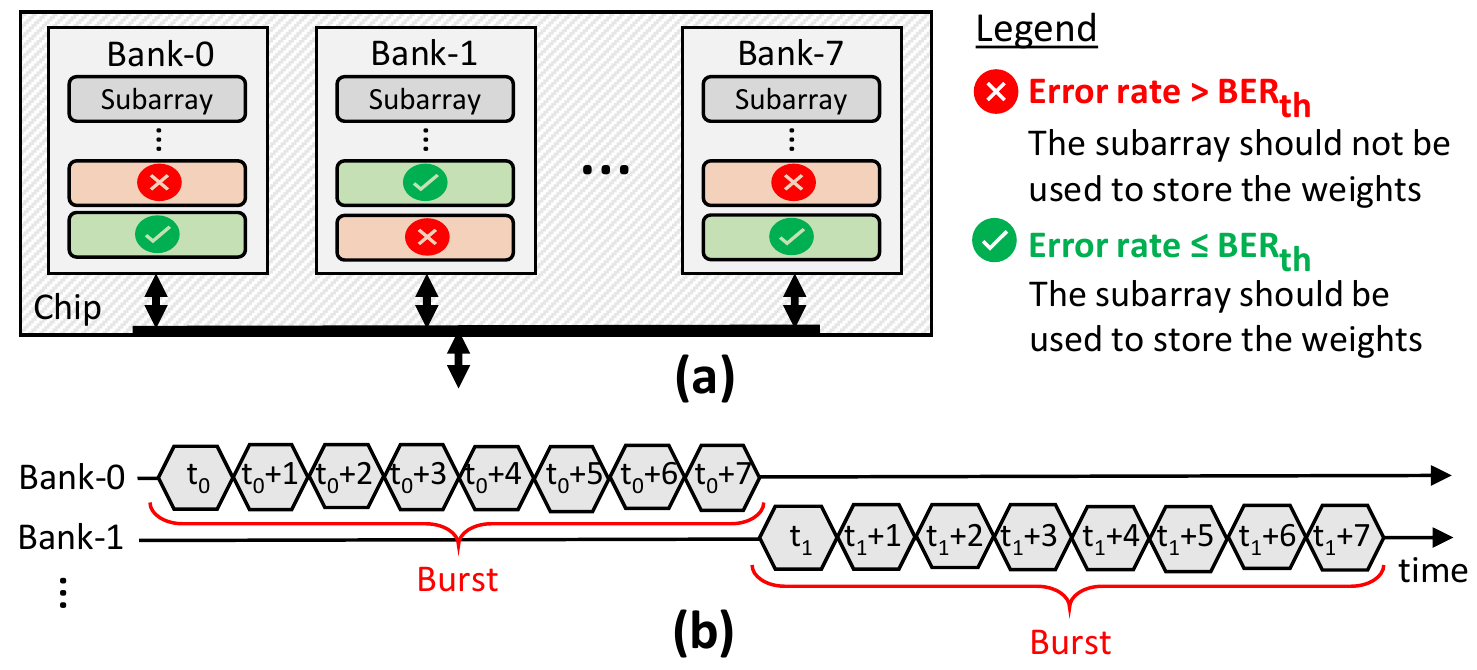}
\vspace{-0.3cm}
\caption{(a) Overview of the proposed DRAM mapping with subarray-level granularity. (b) The DRAM multi-bank burst feature.}
\label{Fig_DRAMmap}
\vspace{-0.2cm}
\end{figure}

The improved SNN model needs to be placed properly in DRAM to ensure that the trained weights are minimally affected by bit errors in DRAM, and hence the classification accuracy is maintained.
Towards this, \textit{our SparkXD employs a DRAM mapping policy to properly map the SNN weights in DRAM, while optimizing the DRAM energy-per-access.}
The proposed DRAM mapping is illustrated in Fig.~\ref{Fig_DRAMmap}(a), and following are the key ideas.
\begin{itemize}[leftmargin=*]
    \item The weights are mapped in the appropriate DRAM partition (e.g., channel, rank, chip, bank, or subarray), whose error profile meets the BER requirement, i.e., error rate $\leq$ maximum tolerable BER ($BER_{th}$).   
    In this work, we consider the subarray-level granularity for our DRAM mapping, since such a granularity level allows us to exploit the following aspects. 
    \begin{itemize}
        \item The DRAM multi-bank burst feature, which is available in the commodity DRAMs, can be employed to increase the data throughput.  
        Its timing diagram is illustrated in Fig.~\ref{Fig_DRAMmap}(b).
        \item The subarray-level parallelism, which is available in new DRAM architectures (such as in \cite{Ref_Putra_DRMap_DAC20}), can also be employed to increase the data throughput.
    \end{itemize}   
    \item The weights are mapped in a way to maximize row buffer hits and minimize row buffer conflicts, for optimizing the DRAM energy-per-access. 
    The reason is that the row buffer hit incurs the least DRAM access energy than other access conditions, as suggested by our experiments in Fig.~\ref{Fig_CaseStudy}(b).
\end{itemize}

To effectively implement the above ideas, we devise Algorithm~\ref{Alg_DRAMmapping} with the following key steps.
(1) Identify the subarrays whose error rate $\leq$ $BER_{th}$. We refer them to as \textit{safe subarrays} which should be prioritized for storing the weights (shown in Algorithm~\ref{Alg_DRAMmapping}, line 7). 
(2) To maximize the row buffer hits and exploit the multi-bank burst feature, the DRAM mapping sequence in each DRAM chip follows the following policy (shown in Algorithm~\ref{Alg_DRAMmapping}, lines 3-8).
\begin{itemize}[leftmargin=*]
   \item \textbf{Step-\circled{1}}: 
   We prioritize to map the data in different columns of the same row, to maximize row buffer hits. 
   If all columns in the same row are filled, then the remaining data is mapped to a target subarray in a different bank. 
   \item \textbf{Step-\circled{2}}: 
   If the target subarray in the target bank meets the BER requirement, the remaining data is mapped to this subarray like in \textbf{Step-\circled{1}}. 
   Otherwise, this subarray is not used and we move the target to a subarray in a different bank. 
   Afterwards, we perform the \textbf{Step-\circled{2}} again.
   If all columns in the same row of all banks are filled/unavailable or unsafe, then the remaining data is mapped to a different subarray in the target bank.
   \item \textbf{Step-\circled{3}}: 
   In the target subarray, remaining data is mapped in the same fashion as in \textbf{Steps-\circled{1}} and \textbf{\circled{2}}. 
   If all columns in the same row of all safe subarrays across all banks are filled, then the remaining data is mapped to a different row, and we perform  \textbf{Steps-\circled{1}} to \textbf{\circled{3}}.
   \item \textbf{Step-\circled{4}}: If some data still  remains, it is mapped to different chips, ranks, and channels respectively, using the \textbf{Steps-\circled{1}} to \textbf{\circled{3}}.
\end{itemize}

\vspace{-0.2cm}
\begin{algorithm}[hbtp]
\scriptsize
\caption{\color{black} The proposed DRAM mapping}
\label{Alg_DRAMmapping}
\begin{algorithmic}[1]
\renewcommand{\algorithmicrequire}{\textbf{INPUT:}}
\renewcommand{\algorithmicensure}{\textbf{OUTPUT:}}
\REQUIRE \textbf{(1)} DRAM ($DRAM$), number of channel-per-module ($n_{ch}$), number of rank-per-channel ($n_{ra}$), number of chip-per-rank ($n_{cp}$), number of bank-per-chip ($n_{ba}$), number of subarray-per-bank ($n_{su}$), number of row-per-subarray ($n_{ro}$), number of column-per-row ($n_{co}$);
\textbf{(2)} Error rates of subarrays ($subarray\_rate$);
\textbf{(3)} Data ($data$); \\
\ENSURE DRAM ($DRAM$); \\
\vspace{0.1cm}
\renewcommand{\algorithmicrequire}{\textbf{BEGIN}}
\renewcommand{\algorithmicensure}{\textbf{END}}
\REQUIRE \hspace{0.1cm} \\   
    \textbf{Process}: \\
    \FOR{$ch = 0$ to $(n_{ch}-1)$}
    \FOR{$ra = 0$ to $(n_{ra}-1)$}
    \FOR{$cp = 0$ to $(n_{cp}-1)$}
    \FOR{$ro = 0$ to $(n_{ro}-1)$}
    \FOR{$su = 0$ to $(n_{su}-1)$}
    \FOR{$ba = 0$ to $(n_{ba}-1)$}
      \IF{$subarray\_rate[ch, ra, cp, ba, su] \leq BER_{th}$}
        \FOR{$co = 0$ to $(n_{co}-1)$} 
          \STATE $DRAM[ch, ra, cp, ba, su, ro, co] \leftarrow  data$;
        \ENDFOR
      \ENDIF
    \ENDFOR
    \ENDFOR
    \ENDFOR
    \ENDFOR
    \ENDFOR
    \ENDFOR  
    \RETURN $DRAM$;
\ENSURE 
\end{algorithmic}
\end{algorithm}
\setlength{\textfloatsep}{2pt}

\vspace{-0.4cm}
\section{Evaluation Methodology}
\label{Sec_EvalMethod}
\vspace{-0.1cm}

\begin{figure}[hbtp]
\vspace{-0.3cm}
\centering
\includegraphics[width=0.9\linewidth]{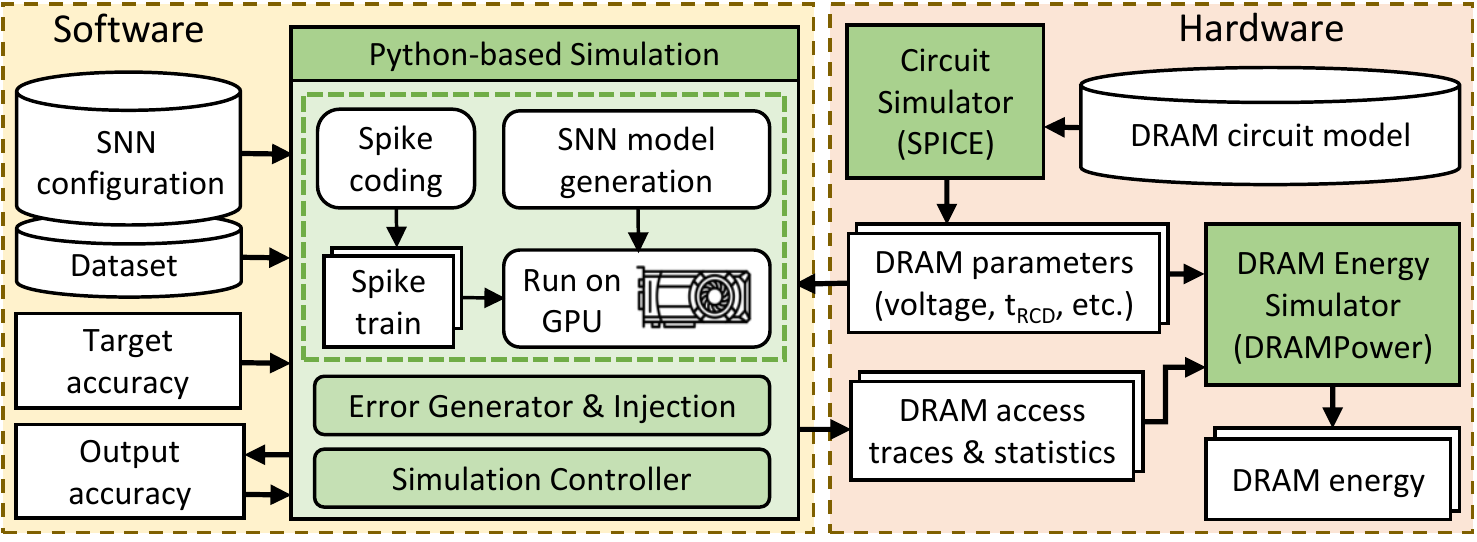}
\vspace{-0.3cm}
\caption{Experimental setup and tool flow.}
\label{Fig_ExpSetup}
\vspace{-0.3cm}
\end{figure}

The experimental setup for evaluating our SparkXD framework is shown in Fig.~\ref{Fig_ExpSetup}. Following are the details of the experimental setup. 

\textbf{Accuracy Evaluation:} 
We use a Python-based simulation \cite{Ref_Hazan_BindsNET_FNINF18} with FP32 precision that run on GPGPU (Nvidia RTX 2080 Ti) and Embedded GPU (Nvidia Jetson Nano), to show the generality of our solution across different compute and memory capabilities.
We use the MNIST and the Fashion MNIST, and employ the rate coding and the Poisson distribution for converting the input samples into spike trains. 
For comparison partner, we consider the baseline SNN model which is trained without considering DRAM errors. 
We also define the target accuracy to be within 1\% of the baseline SNN with accurate DRAM.
For network architecture, we employ the fully-connected SNN with different number of neurons: 400, 900, 1600, 2500, and 3600 (i.e., N400, N900, N1600, N2500, and N3600, respectively).

\textbf{Error Generation and Injection:} 
We generate bit errors based on the DRAM error model-0, and inject them into the locations in DRAM. 
Based on the DRAM mapping policy, the data bits that are stored in the locations with errors, will be flipped. 
For the baseline mapping, we place the weights in subsequent addresses in a DRAM bank. 
For the mapping in SparkXD, we use the proposed Algorithm~\ref{Alg_DRAMmapping}. 

\textbf{DRAM Energy Evaluation:} 
We employ the DRAM circuit model from \cite{Ref_Chang_Voltron_POMACS17} and the SPICE simulator to extract the DRAM parameters (e.g., $V_{supply}$, $t_{RCD}$, $t_{RAS}$, $t_{RP}$),  
while considering LPDDR3-1600 4Gb DRAM configuration which is representative for the main memory of energy-constrained embedded systems. 
The accurate DRAM is operated with 1.35V of $V_{supply}$, while the approximate one is operated within the range of 1.025V-1.325V. 
Afterwards, we use the state-of-the-art and cycle-accurate DRAMPower \cite{Ref_DRAMpower} that consider the extracted DRAM parameters, as well as the DRAM access traces and statistics for estimating the DRAM access energy.

\vspace{-0.1cm}
\section{Results and Discussions}
\label{Sec_Results}
\vspace{-0.1cm}

\begin{figure*}[hbtp]
\centering
\includegraphics[width=0.93\linewidth]{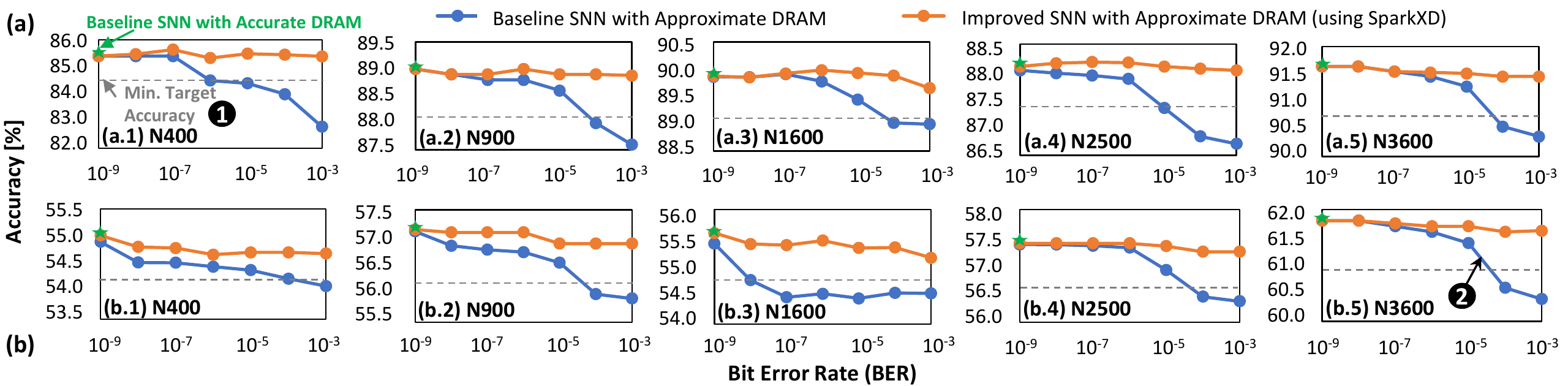}
\vspace{-0.3cm}
\caption{The accuracy of the baseline SNN with accurate DRAM, the baseline SNN with approximate DRAM, and the improved SNN with approximate DRAM, for (a) the MNIST and (b) the Fashion MNIST datasets, across different BER values and different network sizes.} 
\label{Fig_Results_Accuracy}
\vspace{-0.1cm}
\end{figure*}

\begin{figure*}[t]
\vspace{-0.1cm}
\centering
\includegraphics[width=0.93\linewidth]{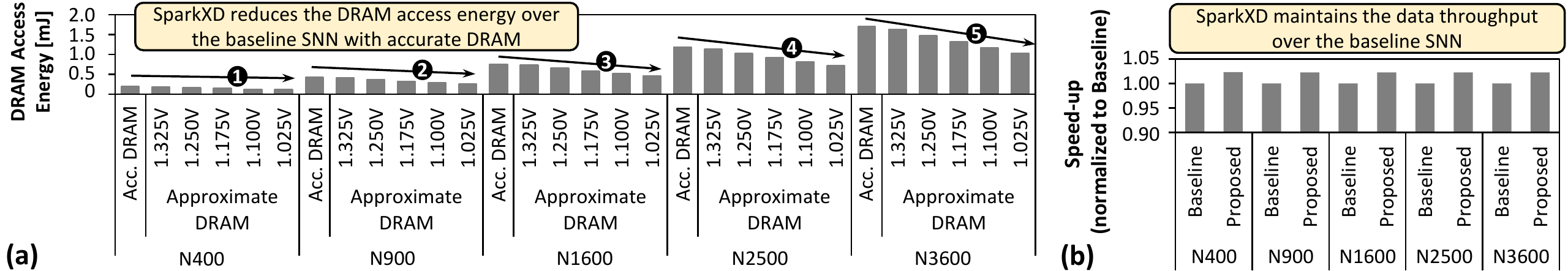}
\vspace{-0.3cm}
\caption{(a) The DRAM energy in an inference incurred by the baseline SNN with accurate DRAM and the SparkXD-based improved SNN with approximate DRAM, across different $V_{supply}$ values, network sizes, and datasets/workloads. (b) The speed-up achieved by the SparkXD over the baseline SNN.} 
\label{Fig_Results_DRAMenergySpeed}
\vspace{-0.5cm}
\end{figure*}

\subsection{Improvements of the SNN Error Tolerance}
\label{Sec_Results_ErrorTolerance}
\vspace{-0.1cm}

Fig.~\ref{Fig_Results_Accuracy} presents the results of accuracy for the baseline SNN with accurate DRAM, the baseline SNN with approximate DRAM, and the improved SNN with approximate DRAM. 

We observe that the baseline SNN with approximate DRAM suffers from the accuracy degradation, as compared to the baseline SNN with accurate DRAM. Here, the accuracy decreases as the error rate increases. 
The reason is that, the weight bits are corrupted (flipped) when they are stored in the approximate DRAM, and these weights are not trained to adapt with such bit flips. 
On the other hand, the improved SNN with approximate DRAM, can maintain the accuracy within 1\% of the baseline SNN with accurate DRAM, across different error rates, different network sizes, and different datasets. 
The minimum target accuracy is shown by a dashed-line pointed by label-\circledB{1} in Fig.~\ref{Fig_Results_Accuracy}.
The reason is that, the SparkXD incorporates the error profile from the approximate DRAM in the training process. 
In this manner, \textit{our SparkXD trains the weights to adapt to the presence of bit errors, and thereby improving the SNN error tolerance.}

Furthermore, we observe the impact of bit error locations in DRAM.
The locations of bit errors are distributed based on the locations of weak cells and generated from the Error Model-0. 
Here, we found that when the bit errors flip the most significant bits (MSBs) of weights, then they change the corresponding weight values and the accuracy may be decreased significantly, as shown by label-\circledB{2} in Figs.~\ref{Fig_Results_Accuracy}. 
On the other hand, when the bit errors flip the less significant bits of weights, then the errors may not significantly change the corresponding weight values and the accuracy is not much affected. 
Therefore, in general, how the bit errors are distributed in DRAM has impact on the accuracy of SNN inference. 

\vspace{-0.1cm}
\subsection{DRAM Access Energy Savings}
\label{Sec_Results_Energy}
\vspace{-0.1cm}

Fig.~\ref{Fig_Results_DRAMenergySpeed}(a) presents the experimental results of the DRAM access energy for the baseline SNN with accurate DRAM and the improved SNN with approximate DRAM, across different $V_{supply}$ values, different network sizes, and different datasets/workloads (the MNIST and the Fashion MNIST). 
These two workloads have similar DRAM access energy, since they have the similar number of weights and number of DRAM accesses for inference. 
The reduction of the $V_{supply}$ to 1.325V, 1.25V, 1.175V, 1.1V, and 1.025V using our SparkXD, saves the DRAM energy by the 3.84\%, 13.33\%, 22.69\%, 31.12\%, 39.46\% on average respectively, across different network sizes.
For each network size, following are the detailed results when reducing the $V_{supply}$ to 1.325V, 1.25V, 1.175V, 1.1V, and 1.025V. 
\begin{itemize}[leftmargin=*]
    \item \textit{Label}-\circledB{1}: For N400,
    the DRAM access energy is saved by 3.81\%, 13.31\%, 22.66\%, 31.09\%, and 39.44\%, respectively.
    \item \textit{Label}-\circledB{2}: For N900, 
    the DRAM access energy is saved by the 3.84\%, 13.33\%, 22.69\%, 31.12\%, and 39.46\%, respectively.
    \item \textit{Label}-\circledB{3}: For N1600, 
    the DRAM access energy is saved by the 3.85\%, 13.34\%, 22.69\%, 31.12\%, 39.46\% respectively.
    \item \textit{Label}-\circledB{4}: For N2500, 
    the DRAM access energy is saved by the 3.85\%, 13.34\%, 22.69\%, 31.12\%, 39.47\% respectively.
    \item \textit{Label}-\circledB{5}: For N3600, 
    the DRAM access energy is saved by the 3.85\%, 13.34\%, 22.69\%, 31.12\%, 39.47\% respectively.
\end{itemize}

These results show that \textit{SparkXD substantially reduces the DRAM access energy, compared to the baseline SNN with accurate DRAM.}
The reasons are two folds: (1) the reduced supply voltage makes the DRAM operates at the reduced operational parameters (e.g., voltage, power), and (2) the proposed DRAM mapping maximizes the row buffer hits and optimizes the DRAM energy-per-access (see Table~\ref{Table_EnergyPerAccess}). 
Furthermore, we also observe that the SparkXD also maintains the throughput as compared to the baseline SNN across different network sizes (i.e., it achieves 1.02x speed-up on average), as shown in Fig.~\ref{Fig_Results_DRAMenergySpeed}(b).  
The reason is that, the proposed DRAM mapping maximizes the row buffer hits and exploit the DRAM multi-bank burst feature.  
In this manner, \textit{our SparkXD improves the DRAM access energy, and thereby improving the overall energy-efficiency of SNN inference, while maintaining the data throughput performance}. 

\vspace{-0.1cm}
\begin{table}[hbtp]
  	\vspace{-0.2cm}
    \caption{Energy Savings over the Baseline SNN with Accurate DRAM Considering the DRAM Energy-per-Access.}
	\label{Table_EnergyPerAccess}
	\vspace{-0.2cm}
	\centering
	\scriptsize
	\begin{tabular}{|c|c|c|c|c|c|}
		\hline
		\textbf{Type of Energy Saving} & \textbf{1.325V} & \textbf{1.250V} & \textbf{1.175V} & \textbf{1.100V} & \textbf{1.025V} \\
		\hline
		\hline
		DRAM energy-per-access & 3.92\% & 14.29\% & 24.33\% & 33.59\% & 42.40\% \\
		\hline
	\end{tabular}
    \vspace{-0.2cm}
\end{table}

\vspace{-0.2cm}
\section{Conclusion}
\label{Sec_Conclusion}
\vspace{-0.1cm}

We propose a novel SparkXD framework to achieve resilient and energy-efficient SNN inference under approximate DRAM, through error-aware SNN training, SNN error-tolerance analysis, and error-aware DRAM mapping.
SparkXD reduces the DRAM energy by ca. 40\% on average, 
while maintaining the accuracy within 1\% of the baseline SNN with accurate DRAM. 
Furthermore, our work would enable further studies on the resilient and energy-efficient SNN. 

\vspace{-0.05cm}


\section{Acknowledgment}
\vspace{-0.05cm}
This work was partly supported by Intel Corporation through Gift funding for the project ”Cost-Effective Dependability for Deep Neural Networks and Spiking Neural Networks”, and by the Indonesia Endowment Fund for Education (IEFE/LPDP) Scholarship Program from Ministry of Finance, Indonesia.

\vspace{-0.05cm}
\bibliographystyle{IEEEtran}
\bibliography{bibliography}

\begin{thebibliography}{10}
\providecommand{\url}[1]{#1}
\csname url@samestyle\endcsname
\providecommand{\newblock}{\relax}
\providecommand{\bibinfo}[2]{#2}
\providecommand{\BIBentrySTDinterwordspacing}{\spaceskip=0pt\relax}
\providecommand{\BIBentryALTinterwordstretchfactor}{4}
\providecommand{\BIBentryALTinterwordspacing}{\spaceskip=\fontdimen2\font plus
\BIBentryALTinterwordstretchfactor\fontdimen3\font minus
  \fontdimen4\font\relax}
\providecommand{\BIBforeignlanguage}[2]{{%
\expandafter\ifx\csname l@#1\endcsname\relax
\typeout{** WARNING: IEEEtran.bst: No hyphenation pattern has been}%
\typeout{** loaded for the language `#1'. Using the pattern for}%
\typeout{** the default language instead.}%
\else
\language=\csname l@#1\endcsname
\fi
#2}}
\providecommand{\BIBdecl}{\relax}
\BIBdecl

\bibitem{Ref_Pfeiffer_DLSNN_FNINS18}
M.~Pfeiffer and T.~Pfeil, ``Deep learning with spiking neurons: Opportunities
  and challenges,'' \emph{Frontiers in Neuroscience}, vol.~12, 2018.

\bibitem{Ref_Akopyan_TrueNorth_TCAD15}
F.~{Akopyan} \emph{et~al.}, ``Truenorth: Design and tool flow of a 65 mw 1
  million neuron programmable neurosynaptic chip,'' \emph{IEEE TCAD}, vol.~34,
  2015.

\bibitem{Ref_Roy_PEASE_ISLPED17}
A.~{Roy} \emph{et~al.}, ``A programmable event-driven architecture for
  evaluating spiking neural networks,'' in \emph{Proc. of ISLPED}, July 2017,
  pp. 1--6.

\bibitem{Ref_Sen_ApproxSNN_DATE17}
S.~{Sen} \emph{et~al.}, ``Approximate computing for spiking neural networks,''
  in \emph{Proc. of DATE}, March 2017, pp. 193--198.

\bibitem{Ref_Krithivasan_SpikeBundle_ISLPED19}
S.~{Krithivasan} \emph{et~al.}, ``Dynamic spike bundling for energy-efficient
  spiking neural networks,'' in \emph{Proc. of ISLPED}, July 2019, pp. 1--6.

\bibitem{Ref_Rathi_PruneQuantizeSNN_TCAD18}
N.~{Rathi} \emph{et~al.}, ``Stdp-based pruning of connections and weight
  quantization in spiking neural networks for energy-efficient recognition,''
  \emph{IEEE TCAD}, vol.~38, no.~4, pp. 668--677, April 2019.

\bibitem{Ref_Putra_FSpiNN_TCAD20}
R.~V.~W. {Putra} and M.~{Shafique}, ``Fspinn: An optimization framework for
  memory-efficient and energy-efficient spiking neural networks,'' \emph{IEEE
  TCAD}, vol.~39, no.~11, pp. 3601--3613, 2020.

\bibitem{Ref_DRAMpower}
\BIBentryALTinterwordspacing
K.~Chandrasekar \emph{et~al.} Drampower. [Online]. Available: \url{http://www.
  drampower.info}
\BIBentrySTDinterwordspacing

\bibitem{Ref_Chandrasekar_DRAMPower}
K.~Chandrasekar, ``High-level power estimation and optimization of drams,''
  Ph.D. dissertation, TU Delft, 2014.

\bibitem{Ref_Chang_Voltron_POMACS17}
K.~K. Chang \emph{et~al.}, ``Understanding reduced-voltage operation in modern
  dram devices: Experimental characterization, analysis, and mechanisms,''
  \emph{ACM POMACS}, vol.~1, no.~1, Jun. 2017.

\bibitem{Ref_Gautrais_SpikeCoding_Bio98}
J.~Gautrais and S.~Thorpe, ``Rate coding versus temporal order coding: a
  theoretical approach,'' \emph{Biosystems}, vol.~48, no.~1, pp. 57--65, 1998.

\bibitem{Ref_Thorpe_RankOrder_Springer98}
S.~Thorpe and J.~Gautrais, ``Rank order coding,'' in \emph{Computational
  neuroscience}.\hskip 1em plus 0.5em minus 0.4em\relax Springer, 1998, pp.
  113--118.

\bibitem{Ref_Park_BurstSNN_DAC19}
S.~Park \emph{et~al.}, ``Fast and efficient information transmission with burst
  spikes in deep spiking neural networks,'' in \emph{Proc. of DAC}, 2019,
  p.~53.

\bibitem{Ref_Putra_DRMap_DAC20}
R.~V.~W. {Putra} \emph{et~al.}, ``Drmap: A generic dram data mapping policy for
  energy-efficient processing of convolutional neural networks,'' in
  \emph{Proc. of DAC}, 2020, pp. 1--6.

\bibitem{Ref_Koppula_EDEN_MICRO19}
S.~Koppula \emph{et~al.}, ``Eden: Enabling energy-efficient, high-performance
  deep neural network inference using approximate dram,'' in \emph{Proc. of
  MICRO}, 2019, p. 166–181.

\bibitem{Ref_Hazan_BindsNET_FNINF18}
H.~Hazan \emph{et~al.}, ``Bindsnet: A machine learning-oriented spiking neural
  networks library in python,'' \emph{Frontiers in Neuroinformatics}, 2018.

\end{thebibliography}

\end{document}